\documentclass{ws-p8-50x6-00}
\def\beq{\begin{equation}}
\def\eeq{\end{equation}}
\def\be{\begin{eqnarray}}
\def\ee{\end{eqnarray}}
\def\ci{\cite}
            
\def\lsim{\mathrel{\rlap{\lower4pt\hbox{\hskip1pt$\sim$}}
    \raise1pt\hbox{$<$}}}         
\def\gsim{\mathrel{\rlap{\lower4pt\hbox{\hskip1pt$\sim$}}
    \raise1pt\hbox{$>$}}}         

\begin{document}

\title{Is the Equation of State of strongly interacting matter 
observable ?}

\author{Omar Benhar}

\address{INFN, Sezione di Roma \\
Dipartimento di Fisica, Universit\`a ``La Sapienza'\\
I-00161 Roma, Italy \\
E-mail: benhar@roma1.infn.it}

\maketitle

\abstracts{
I review the available empirical information on the equation of state 
of cold strongly interacting matter, as well as the prospects for
obtaining new insights from the experimental study of gravitational 
waves emitted by neutron stars. 
}

\section{Introduction}

The equation of state (EOS) is a nontrivial relation linking the 
thermodynamic variables specifying the state of a physical system \ci{huang}. 
The best known example of EOS is Boyle's {\it ideal  gas law}, stating
 that the pressure of a collection of $N$ noninteracting, pointlike 
classical particles, enclosed in a volume $\Omega$, grows linearly with 
the temperature $T$ and the average particle density $n = N/\Omega$. 

The ideal gas law provides a good description of very dilute systems. 
In general, the EOS can be written as an expansion of the pressure, $P$, 
in powers of the density (we use units such that Boltzmann's 
constant is $K_B = 1$):
\beq
P = nT\ \left[ 1 + n B(T)  + n^2 C(T) + \ldots \right]\ .
\label{virial}
\eeq
The coefficients appearing in the above series, that goes under the name of 
{\it virial expansion}, are functions of temperature only. They describe 
the deviations from the ideal gas law and can be calculated in terms of 
the underlying elementary interaction. Therefore, the EOS carries a 
great deal of dynamical 
information, and its knowledge makes it possible to establish a 
link between measurable {\it macroscopic} quantities, such as 
pressure or temperature, and the forces acting 
between the constituents of the system at {\it microscopic} level. 

\begin{figure}[ht]
\begin{center}
\epsfxsize=17pc 
\epsfbox{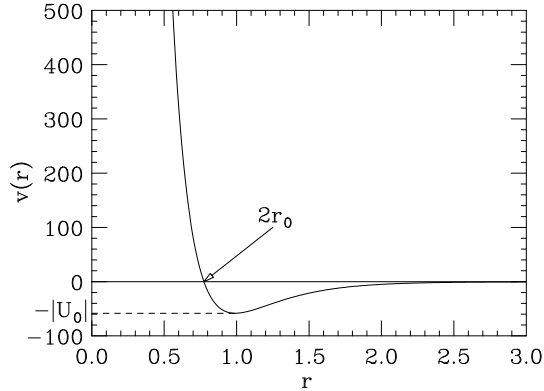} 
\caption{Behavior of the potential describing the interactions
between constituents of a van der Waals fluid (the interparticle 
distance $r$ and $v(r)$ are both given in arbitrary units). \label{vdWpot}}
\end{center}
\end{figure}

This point is best illustrated by the van der Waals EOS, 
which describes a collection of particles interacting through
a potential featuring a strong repulsive core followed by a weaker
attactive tail (see fig.\ref{vdWpot}). At $|U_0|/T << 1$, $U_0$ being 
the strength of the attractive part of the potential, 
the van der Waals EOS takes the simple form 
\beq
P = \frac{nT}{1 - nb} - a n^2\ ,
\eeq
and the two quantities $a$ and $b$, taking into account interaction 
effects, can be directly related to the potential $v(r)$ 
through
\beq
a = \pi \int_{{2r_0}}^\infty |v(r)|^2\ r^2 dr \ \ \ \ \ , 
 \ \ \ \ \ b = \frac{16}{3}\pi r_0^3 \ ,
\eeq
where $2r_0$ denotes the radius of the repulsive core 
(see fig.\ref{vdWpot}).

In spite of its simplicity, the van der Waals EOS describes most
of the features of both the gas and liquid phases of the system, as well 
as the nature of the phase transition.

\begin{figure}[ht]
\begin{center}
\epsfxsize=17pc
\epsfbox{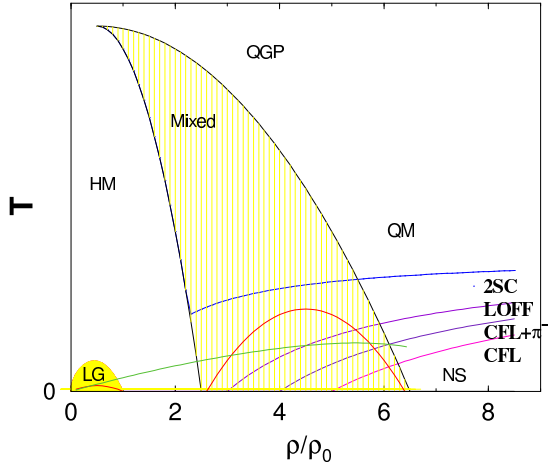}
\caption{Temperature {\it vs} baryon density phase diagram of 
charge neutral strongly interacting matter in $\beta$-equilibrium.
Hatched areas correspond to mixed phases of hadronic matter (HM)
and quark matter (QM/QGP), as well as the nuclear liquid-gas 
(from ref.\protect\ci{phasedref}). \label{phased}}
\end{center}
\end{figure}

The EOS of strongly interacting matter reflects the complexity
of the underlying dynamics, that makes the phase diagram extremely rich.
As an example, fig.\ref{phased} shows the phase diagram of charge 
neutral strongly interacting matter in $\beta$-equilibrium\ci{phasedref}.

In this talk, I will review the available empirical information on the 
EOS of strongly interacting matter at $T \sim 0$, and the extent to which this 
information can be used to constrain theoretical models. The knowledge coming 
from nuclear systematics and neutron stars data
is summarized in Sections 2 and 3, respectively. Section 4 provides
a short overview of the theoretical models of neutron star matter at 
supranuclear density, while Section 5 is devoted to the prospects 
for obtaining new insights from the experimental study of gravitational 
waves emitted by neutron stars. Finally, Section 6 summarizes the main 
results and states the conclusions.  

\section{Constraints on the EOS at $T$=0 from nuclear data}

Under standard terrestrial conditions, strongly interacting matter is
observed in form of rather small, cold chunks: the atomic nuclei. 

Nuclei are self-bound systems consisting of $Z$ protons and $(A-Z)$ neutrons. 
Their smallness is a consequence of the electrostatic repulsion between 
protons, limiting $A$ to $\sim$ 200, while the fact that they can be described 
as cold objects follows from the observation that thermal energies are 
negligible in comparison to the large proton and neutron Fermi energies.

The body of data on nuclear masses provides a constraint on the 
density dependence of the energy per particle $e = E/A$ at zero temperature, 
related to the EOS through
\beq
P(n,T=0) = - \left( \frac {\partial E}{\partial \Omega} \right)_{T=0} = 
n^2  \left( \frac{de}{dn} \right)_{T=0}\ .
\eeq

The $A$-dependence of the nuclear binding energy (i.e. the difference
between the measured nuclear mass and the sum of the constituent masses) 
is well described by the semiempirical formula
\beq
B({\rm Z},{\rm A}) = a_{\rm V} {\rm A} + a_{\rm s} {\rm A}^{2/3} 
+ a_{\rm c} \frac{{\rm Z}^2}{{\rm A}^{1/3}} 
+ a_{\rm A} \frac{ ({\rm A}-2{\rm Z})^2 }{ 4 {\rm A} }
- \lambda \  a_{\rm p} \frac{1}{{\rm A}^{1/2}}\ ,
\label{mass:f}
\eeq
whose terms account for nuclear interactions, taking
place both in the interior of the nucleus and on its surface, electrostatic
interactions between protons, isospin asymmetry and shell effects.

The coefficient of the term linear in $A$ yields the binding energy 
of {\it symmetric nuclear matter}, an ideal uniform system consisting of
equal number of protons and neutrons coupled by strong 
interactions only. The equilibrium density of such a system, $n_0$, can be 
inferred exploiting saturation of nuclear densities, i.e. the fact
that the central density of atomic nuclei, measured by elastic 
electron-nucleus scattering, does not depend upon $A$ for 
large $A$. 

The empirical equilibrium properties of nuclear matter are
\beq
e_0 = e(n=n_0,T=0) = - 16\ {\rm MeV}/A \ \ , 
\ \ n_0 \sim .16\  {\rm fm}^{-3}\ .
\eeq
In the vicinity of the equilibrium density $e(n,T=0)$ can be expanded 
according to (as we will only discuss the EOS at $T=0$, 
the dependence upon $T$ will be omitted hereafter)
\beq
e(n) \approx e_0 + \frac{1}{2}\ \frac{K}{9}\ \frac{(n-n_0)^2}{n_0^2}\ , 
\label{quad}
\eeq
where 
\beq
K = 9\ n_0^2 \left( \frac{\partial^2 e}{\partial n^2} \right)_{n=n_0} =
9  \left( \frac{\partial P}{\partial n} \right)_{n=n_0}                   
\eeq
is the (in)compressibility module, that can be extracted from the 
measured excitation energies of nuclear vibrational states. Empirical
values range from $K \sim 200$ MeV (corresponding to more compressible 
nuclear matter, i.e. to a {\it soft} EOS) to $K \sim 300$ MeV (corresponding
to a {\it stiff} EOS) \ci{compress}. 

It has to be emphasized that the quadratic extrapolation of Eq.(\ref{quad})
cannot be expected to work far from equilibrium density. In fact, assuming 
a parabolic behavior of $e(n)$ at large $n$ ($>> n_0$) leads to 
predict a speed of sound in matter, $v_s$, larger than the speed of light, i.e.
\beq
\left( \frac{v_s}{c} \right) = \frac{1}{n} 
\left( \frac{\partial P}{\partial e} \right) > 1\ ,
\label{speed}
\eeq
regardless of the value of $K$. 

Eq.(\ref{speed}) shows that causality requires
\beq
\left( \frac{\partial P}{\partial \epsilon} \right) < 1\ ,
\label{speed2}
\eeq
where $\epsilon = E/\Omega$ is the energy-density. For a noniteracting 
Fermi gas $\epsilon \propto n^{4/3}$, implying 
(the equal sign corresponds to massless fermions)
\beq
P \leq \frac{\epsilon}{3} \ \ , \ \ \left( \frac{v_s}{c} \right)
\leq \frac{1}{3}\ .   
\eeq
In presence of interactions the above limits can be easily exceeded. 
For example, modeling the repulsion between nucleons in terms of a rigid 
core leads to predict infinite pressure at finite density. 

The stiffest EOS compatible with causality is $P = \epsilon$, corresponding 
to $(v_s/c) = 1$. Back in the early 60's Zel'dovich was able to show that
the $v_s=c$ limit, corresponding to $\epsilon \propto n^{2}$, is indeed 
attained in a simple semirealistic theory, in which nucleons are assumed to 
interact through exchange of a vector meson \ci{zel}.

\section{Constraints on the EOS at $T$=0 from neutron stars data}

At very large $A$ ( $\sim 10^{57}$ ), gravity becomes strong enough
to balance the repulsive interactions between nucleons, and nuclear 
matter can occupy macroscopic regions of space. This is the situation 
occurring in the interior of neutron stars, compact astrophysical objects  
(typical values of mass and radius are $\sim$ 1.4 M$_\odot$ and 
$\sim$ 10 Km, respectively) whose central density largely exceeds $n_0$.
\begin{figure}[ht]
\begin{center}
\epsfxsize=18pc
\epsfbox{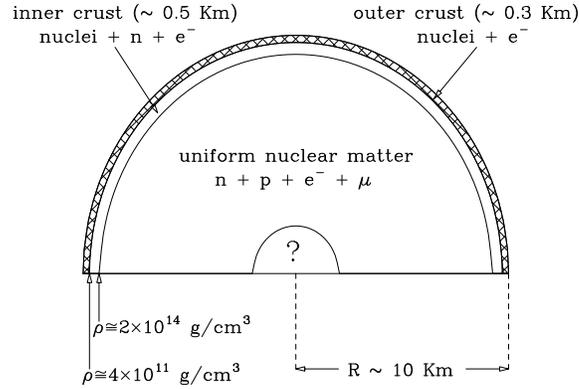}
\caption{Cross section of a neutron star. Note that the equilibrium 
density of nuclear matter corresponds to $\sim$ 2.7 $\times$ 10$^{14}$ 
g/cm$^3$. \label{nsxsec}}
\end{center}
\end{figure}
The structure of a neutron star is schematically represented in 
fig.\ref{nsxsec}. Note that the inner and outer crusts contain a 
comparatively small amount of matter, most of the star mass being 
concentrated in the region of supranuclear densities ($n > n_0$). In addition, 
as all relevant Fermi energies are much larger than the typical temperature
($\lsim$ 10$^9$ K$^\circ$ $\sim$ 100 KeV), neutron stars can be regarded 
as cold object. The main features of the theoretical models of neutron star 
matter will be summarized in Section 4.


Given the EOS describig matter in the interior of  a neutron star, its
mass and radius can be obtained from the Tolman-Oppenheimer-Volkov (TOV)
equations
\beq
\frac{dP(r)}{dr} = - G\
\frac{ \left[ \epsilon(r) + P(r)/c^2 \right]
 \left[ M(r) + 4 \pi r^2 P(r)/c^2 \right] }
{ r^2 \left[ 1 - 2 G M(r) /r c^2 \right] } \ .  
\label{TOV1}
\eeq
\beq
M(r) = 4 \pi \int_0^r {r^\prime}^2 d{r^\prime}
\epsilon({r^\prime}) \ ,    
\label{TOV2}
\eeq
In the above equations, combining hydrostatic equilibrium with Einstein's 
equations of general relativity for a nonrotating star, $P$ and $\epsilon$ 
denote pressure and energy-density, respectively, and $G$ is the 
gravitational constant. 

For any value of the central energy-density $\epsilon_c = \epsilon(r=0)$
Eq.(\ref{TOV1}) can be integrated outward until the star 
radius $R$, defined by the condition $P(R) = 0$, is reached. 
The star mass $M(R)$ can then be obtained from Eq.(\ref{TOV2}).
\begin{figure}[ht]
\begin{center}
\epsfxsize=16pc
\epsfbox{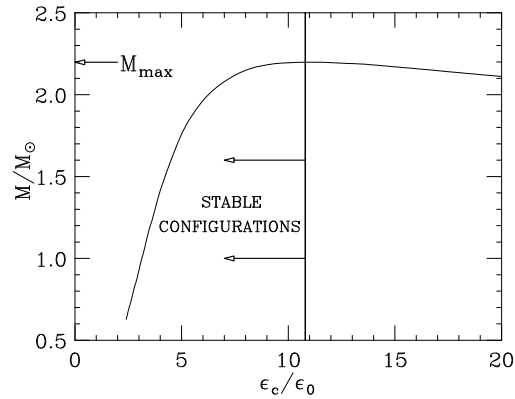}
\caption{Typical dependence of the neutron star mass (in units of
the solar mass $M_\odot$) upon the energy-density at the center of
the star (in units of the energy-density of symmetric nuclear matter at
equilibrium). \label{mvseps}}
\end{center}
\end{figure}

Fig.\ref{mvseps} shows the typical behavior of the neutron star mass
as a function of the central energy-density. Denoting by 
$\overline{\epsilon}$ the central energy-density corresponding to 
the maximum mass, all stable configurations 
have $\epsilon_c < \overline{\epsilon}_c$. 

The value of the maximum mass is mostly determined by the {\it stiffness} of 
the EOS, more incompressible neutron star matter (i.e. {\it stiffer}
EOS) corresponding to larger $M_{max}$. Threrefore, in principle measurements 
of neutron star masses may provide information on the 
behavior of the EOS at $n>n_0$.

The most precise experimental data, obtained from studies of the timing 
of radio pulsars, yield the average value \ci{nsm1}
\beq
M = 1.35 \pm 0.04\ M_\odot\ ,
\label{can:mass}
\eeq
thus constraining any realistic model of EOS to support a stable star 
with mass $\sim 1.4\ M_\odot$. 

Neutron star masses can also be obtained 
from the analysis of binary systems containing an X-ray pulsar. A recent
determination of the mass of the Vela X-1 pulsar yields $M = $
 1.86 $\pm$ 0.33 $M_\odot$ \ci{nsm2}, suggesting the possibility of 
a mass well above the canonical value of Eq.(\ref{can:mass}). If confirmed, the 
existence of a neutron star with $M > 1.8 M_\odot$ would provide a very
stringent constraint, ruling out theoretical models that predict very 
soft EOS.

\section{Theoretical models of neutron star matter at 
supranuclear density.}

Theoretical models of neutron star matter are based on different 
approaches, involving different degrees of freedom, which are 
expected to be applicable in different density regimes.

In the range $n_0 \lsim n \lsim 4 n_0$ matter in the interior of 
the star is believed to take the form of a cold uniform fluid,
that can be described in terms of hadronic degrees of freedom,  
within the framework of either nonrelativistic many-body theory\cite{pandr} 
or relativistic mean-field approaches\cite{glenb}. 

\begin{figure}[ht]
\begin{center}
\epsfxsize=20pc
\epsfbox{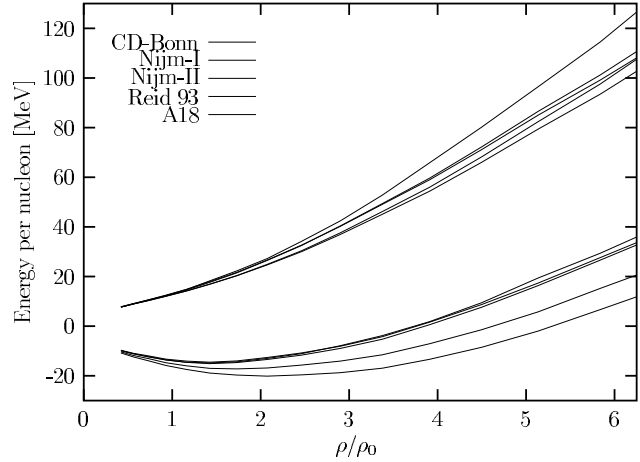}
\caption{Typical density dependence of the energy per nucleon of pure 
neutron matter (upper curves) and symmetric nuclear matter (lower curves), 
obtained from nonrelativistic nuclear many-body 
approaches\protect\cite{pandr}. The density is given in units of 
the equilibrium density of symmetric nuclear matter. \label{e:n}}
\end{center}
\end{figure}

At $n$ just above $n_0$ neutrons dominate, with a 
small fraction of protons and leptons in equilibrium
with respect to the weak interaction processes ($\ell$ denotes either an
electron or a muon)
\beq
n \rightarrow p + \ell + \overline{\nu}_\ell \ \ ,
\ \ p + \ell \rightarrow n + \nu_\ell \ .
\eeq
For any baryon density $n$, minimization of the total energy-density with
the constraints of baryon
number conservation and charge neutrality fixes the relative aboundances
of neutrons, protons and leptons. Note that, as the proton fraction is small, 
typically $< 10\%$, theoretical calculations of the EOS 
involve the study of stronlgly {\it asymmetric} nuclear matter, whose energy 
is most sensitive to the isospin asymmetry term in Eq.(\ref{mass:f}).
The importance of isospin asymmetry is illustrated in fig.\ref{e:n}, showing
the calculated density dependence of
the energy per particle of symmetric nuclear matter ($(Z/A) = 1/2$) and 
pure neutron matter ($(Z/A)=0$). It clearly appears that unlike symmetric 
nuclear matter, whose energy exhibits a minimum, pure neutron matter is 
not self-bound at any density.

As the density increases, different forms of matter, containing 
hadrons other than protons and neutrons, can become energetically 
favored. For example, the weak interaction process
\beq
n + e \rightarrow \Sigma^- + \nu_e\ ,
\eeq
leading to the appearance of a strange baryon, sets in as soon as 
the sum of the electron and neutron chemical potentials exceeds 
$M_{{\Sigma^-}}$ (typically at $n \gsim 2n_0$). At larger density
the production of $\Lambda^0$'s is also energetically allowed. 

\begin{figure}[ht]
\begin{center}
\epsfxsize=22pc
\epsfbox{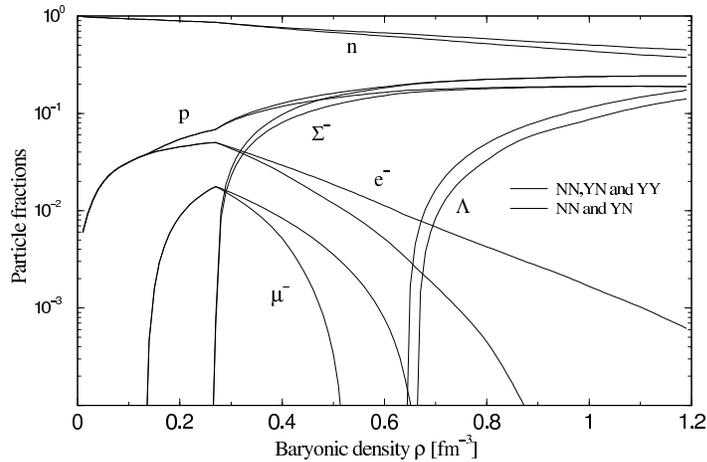}
\caption{Relative aboundances of hadrons and leptons in a hadronic model 
of neutron star matter (taken from ref.\protect\ci{art}). \label{abb}}
\end{center}
\end{figure}

For any fixed baryon density, the relative aboundances of the 
different hadronic and 
leptonic species are dictated by the requirements of equilibrium 
with respect to weak interactions, conservation of baryon number 
and charge neutrality. Fig.\ref{abb} shows the particle fractions
resulting form the nonrelativistic many-body aproach of ref.\ci{art}.

The transition from nucleon matter to hadronic matter is associated 
with a softening of the EOS, as it basically amounts to 
replacing particles carrying large Fermi energies with more dilute, 
and therefore less energetic, strange baryons. Hence, using the EOS of 
hadronic matter as an input for 
the solution of the TOV equations leads to predict a value of the maximum 
mass significantly lower than that obtained using nucleon matter.

At very large density ($n > 4n_0$) a new transition, to a phase in which 
quarks are no longer clustered into nucleons or hadrons, is eventually 
expected to take place. In most theoretical studies of neutron stars 
structure, the EOS of quark 
matter has been estimated using the simple MIT {\it bag model} 
(for a recent review see ref.\ci{HH}). 

The bag model assumes that quarks be confined to a finite region of space 
(the bag), whose volume is limited by a pressure $B$ (the bag constant).
Its simplest implementation includes only massless and noninteracting 
$u$ and $d$ quarks. Equilibrium with respect to the 
process ($\ell$ denotes either an electron or a muon)
\beq
d \rightarrow u + \ell + \overline{\nu}_\ell\ ,
\eeq
requires that the $u$ and $d$ chemical potentials satisfy (note that, 
as the mean free path of neutrinos in dense nucleon matter largely exceeds
their typical radius, neutron stars can be regarded as 
 transparent to neutrinos)
\beq
\mu_d = \mu_u + \mu_\ell \ \ , \ \ \mu_e = \mu_\mu, 
\eeq
whereas the constraint of charge neutrality implies 
\beq
\frac{2}{3} n_u = \frac{1}{3}n_d + n_e + n_\mu\ ,
\eeq
where $n_q$ ($q = u,d$) and $n_\ell$ ($\ell = e,\mu$) denote the quark 
and lepton densities, respectively.

The energy-density reads
\beq
\epsilon = \frac{E}{\Omega} = B 
+ \frac{3}{4\pi^2} \sum_q\ p_{{F_q}}^4\ ,
\label{epsiq}
\eeq
and the quark Fermi momenta are
related to the total baryon density $n$ through
\beq
p_{{F_q}} = (\pi^2 f_q n )^{(1/3)}\ ,
\eeq
$f_q$ being the number of quarks of flavor $q$ per baryon. Finally, 
the pressure can be readily obtained from 
\beq
P = - \left( \frac{\partial E}{\partial \Omega} \right) = - B 
 +  \frac{1}{4\pi^2} \sum_q\ p_{{F_q}}^4\ .
\label{pressq}
\eeq

More realistic models include a massive strange quark and allow 
for one-gluon exchange interactions between quarks of the same 
flavor. 
\begin{figure}[ht]
\begin{center}
\epsfxsize=14pc
\epsfbox{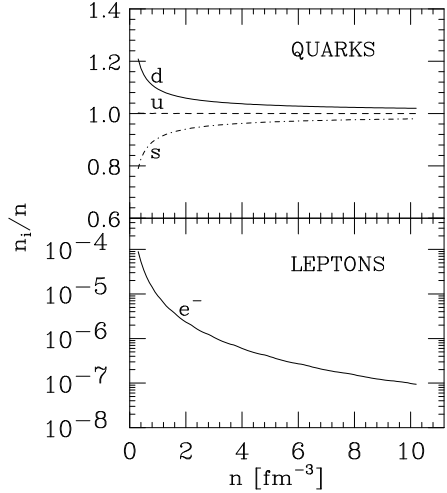}
\caption{Relative aboundances of quarks and leptons in $\beta$-stable
matter consisting of massless $u$ and $d$ quarks and strange quarks 
with mass $m_s = 150$ MeV. All interactions between quarks are 
neglected and the bag constant is set to B = 208 MeV/fm$^3$. 
\label{abbq}}
\end{center}
\end{figure}
At fixed baryon density the quark and lepton fractional densities are 
determined by the requirements of chemical equilibrium, charge
neutrality and baryon number conservation.
Fig.\ref{abbq} shows the results obtained including massless $u$ and
$d$ quarks and strange quarks of mass $m_s$ = 150 MeV, neglecting
all interactions and setting the value of the bag constant to 
$B$ = 208 MeV/fm$^3$. 

In the $n \rightarrow \infty$ limit, the energy per baryon of nucleon 
matter, $e_{NM}$, grows linearly with baryon density, whereas in 
quark matter $e_{QM} \sim n^{1/3}$. Hence, a transition from 
nucleon to quark matter is expected to take place in the inner region
of neutron stars, provided the central density is large enough.

In the simplest implementation, the phase transition is assumed to 
take place at 
constant pressure and chemical potential, so that matter density 
exhibits a discontinuity at the boundary between the two phases. 
According to this picture the star consists of a core of quark matter 
surrounded by nucleon matter. 

The calculations of ref.\ci{panda}, based on nonreativistic 
many-body theory for the description of nucleon matter and the bag model
for the description of quark matter, predict a transition region 
$5.4 < (n/n_0) < 9.8 $ for B = 200 MeV/fm$^3$ and 
$4.9 < (n/n_0) < 7.5 $ for B = 122 MeV/fm$^3$.

The transition to quark matter makes the EOS softer, thus lowering the 
value of the maximum neutron star mass. The authors of ref.\ci{panda}
report a reduction $\Delta M_{max}$ of $\sim$ 10 \% and $\sim$ 20 \% for
B = 200 and 122 MeV/fm$^3$, respectively.

\section{Will detection of gravitational provide new insight ?}

As the pattern of nonradial oscillations of neutron stars 
depends upon the EOS describing matter in the interior of 
the star, detection of gravitational wave emission associated with
the excitation of these modes can in principle provide additional 
constraints on theoretical models of the EOS.

\begin{figure}[ht]
\begin{center}
\epsfxsize=20pc
\epsfbox{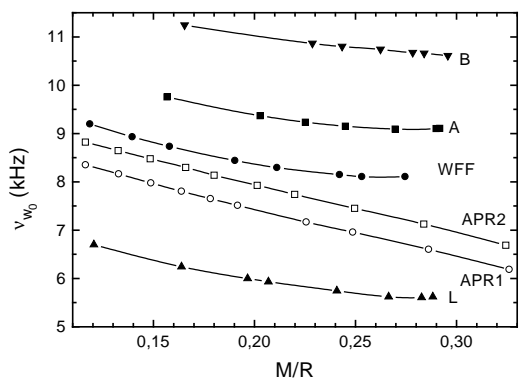}
\caption{Frequencies of the first axial w-mode of a nonrotating
neutron star, plotted as a function of the star compactness ($M/R$).
The different curves correspond to different models of EOS
(taken from ref.\protect\cite{mnras}).  \label{gw1}}
\end{center}
\end{figure}

For example, the complex oscillation frequencies of the axial (i.e. odd
parity) modes of a nonrotating star are eigenvalues of a 
Schr\"odinger-like equation whose potential $V_\ell(r)$ 
explicitely depend upon the EOS according to\cite{mnras}
\beq
V_\ell(r) = \frac{ {\rm e}^{2 \nu(r)} }{r^3}\ \left\{
\ell(\ell+1)r + r^3 \left[ \epsilon(r) - P(r) \right] - 6 M(r) \right\}\ ,
\eeq
where
\beq
\frac{d\nu}{dr} = - \frac{1}{\left[ \epsilon(r) + P(r) \right]}
\frac{dP}{dr}\ .
\eeq
Using realistic EOS one obtains strongly damped eigenmodes, 
called {\it w-modes}, whose frequencies exhibit the behavior diaplayed 
in fig.\ref{gw1}.

The pattern of frequencies strictly reflects the stiffness of the different 
EOS in the relevant density region (typically $1 < (n/n_0) < 5$, softer EOS 
correponding to higher frequencies. For example, the curve labelled B 
has been obtained from a model including nucleons, nucleon excitations 
and strange baryons, predicting a very soft EOS and a maximum mass of 
1.42 $M_\odot$.

It is interesting to note that the dependence upon
the the ratio ($M/R$) is rather weak, so that detection of a given 
frequency may allow one to discriminate between different models of EOS 
(e.g. between model B, corresponding to hadronic matter and model WFF, 
correponding to nucleon matter) regardless of the compactness of the star.

A different class of nonradial oscillations, the {\it g-modes}, are 
associated with the occurrence of a density discontinuity in the 
interior of the star. Early studies of these modes focused on 
discontinuities that are known to be produced by the changes of 
chemical composition in the low density region of the crust, corresponding
to a fractional distance from the surface $(d/R) \lsim$ 10 \%.

As pointed out in Section 4, a different discontinuity may be produced
by the transition from nucleon to quark matter, which is expected to 
take place at much larger density and involve a much larger 
density jump.

It has to be emphasized that, as nucleon and quark matter need 
not to be considered independently charge neutral, in general the 
deconfinement transition may involve the occurrence of a mixed phase 
(e.g. bubbles of quark matter in nucleon matter at lower density or 
{\it viceversa} at larger density) extending over a sizable region of 
space, rather than a sharp separation between the two phases\ci{HH}. 
Whether the mixed phase is energetically favored depends upon the 
balance between the gain in volume energy and the loss associated 
with Coulomb and surface energy\ci{STA}. If Coulomb and surface energies
prevail, a density discontinuity is expected to occur.

\begin{figure}[ht]
\begin{center}
\epsfxsize=18pc
\epsfbox{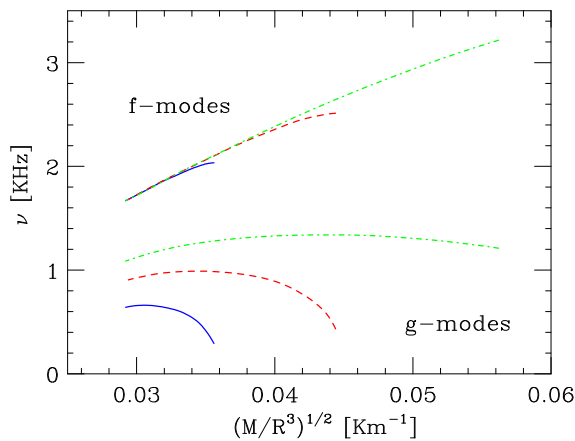}
\caption{Frequencies of the f- and g-mode of a neutron star of mass
$M = 1.4 M_\odot$ as a function of the average density.
Solid, dashed and dot-dash lines correspond to density jumps of
10 \%, 20 \% and 30 \%, respectively. All stellar models have
polytropic exponent $\Gamma = 2$
(see Eq.(\ref{poly}))\protect\cite{giovanni}  \label{gw2}}
\end{center}
\end{figure}
The authors of ref.\ci{giovanni} have recently carried out an exploratory 
calculation
of the frequencies of both the fudamental f-mode and 
the g-mode of a nonrotating 
neutron star, using a simple polytropic EOS with a density discontinuity
$\Delta n$ located at density $n = n_D$:
\begin{equation}
P(n) = \left\{
\begin{array}{lll}
K \left( 1 + \frac{\Delta n}{n_D} \right)^\Gamma n^\Gamma
& \ \ \ \ \ & n < n_D   \\
 & & \\
K n^\Gamma
& \ \ \ \ \ & n > n_D + \Delta n \ .
\end{array}
\right.
\label{poly}
\end{equation}

The results of ref.\ci{giovanni} are summarized in fig.\ref{gw2}, showing
the f- and g-mode frequencies as a function of the average density of the 
star, whose mass is kept fixed at the canonical value of 1.4 $M_\odot$. 
It clearly appears that, while the f-mode frequency exhibits a nearly 
linear growth, largely 
unaffected by the 
abrupt structural change associated with the phase transition, 
the frequencies of the g-mode strongly depend upon the jump
in density $\Delta n$. Hence, a simultanuoeus measurements of the 
frequencies of the two modes would provide information on both 
size and location of the discontinuity.

\section{Conclusions}

\begin{figure}[h]
\begin{center}
\epsfxsize=18pc
\epsfbox{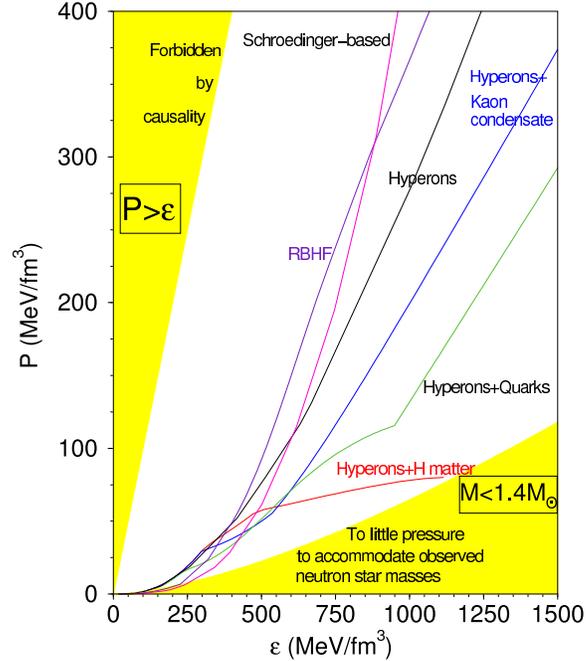}
\caption{EOS of cold $\beta$-stable strongly interacting matter 
resulting from a variety of different theoretical approaches, based on 
both hadronic and quark degrees of freedom. The shaded areas show the region 
forbidden by causality and that corresponding to 
$M_{max} < 1.4\ M_\odot$ (taken from ref.\protect\cite{weber}). \label{EOS0}}
\end{center}
\end{figure}

The extent to which the currently available empirical information 
constrains the EOS of cold $\beta$-stable strongly interacting matter 
is illustrated in fig.\ref{EOS0}.

Although all EOS resulting from nonrelativistic approaches are known to be 
plagued by a noncausal behaviour at very large density, it appears that the 
constraint $P>\epsilon$ is fulfilled by all EOS over a wide range 
of energy-density. 

The requirement that the maximum mass be larger than 
$1.4\ M_\odot$ does not provide a stringent constraint either, as 
all reasonable models support a stable star with $M = 1.4\ M_\odot$. 
However, as pointed out in Section 3, the situation may change should
the mass of the Vela X-1 pulsar be confirmed to be larger than 1.8 $M_\odot$. 
In this case the shaded region in the lower right corner
of fig.\ref{EOS0} would considerably extend upward, thus ruling out models 
that predict very soft EOS. 

The observation of gravitational waves emitted by neutron stars may 
provide valuable new insight, allowing to further constrain theoretical 
models. Detection of gravitational radiation in the relevant few $KHz$ 
frequency range may become possible with the recently proposed gravitational 
laser interferometric detector EURO \cite{EURO}.

As far as theory is concerned, it has to be mentioned that 
the recent progress in Quantum Monte Carlo has made it possible to 
apply this approach to the study of the EOS of pure neutron matter
\cite{stefano}. The extension to the case of $\beta$-stable matter, 
that appears to be feasible, would be of great importance for the 
understanding of neutron star 
matter in the region of not too high density ($n \lsim 2 n_0$), 
where the use of nucleonic degrees of freedom is likely to be 
reasonable.

At larger densities a transition to quark matter 
is expected to take place. A better theoretical description of this
region will require a careful analysis of the nature of the phase 
transition as well as a more realistic description of the deconfined phase.
In this context, a pivotal role is likely to be played by detailed 
quantitative studies of the possible
occurrence of the color superconducting phase predicted by the 
fundamental theory of strong interactions\cite{alford}.


\end{document}